\documentclass{emulateapj}
\newcommand{\alphabcos}{\mbox{cos($\delta$) }}
\newcommand{\testalphabcos}{\mbox{cos($\delta$)}}

\newcommand{\hst}{{\it HST}}

\begin{document}

\shorttitle{Space Motion of NGC 6397}
\shortauthors{Kalirai et al.}

\title{The Space Motion of the Globular Cluster NGC 6397\altaffilmark{1}}

\author{
Jasonjot S. Kalirai\altaffilmark{2,3}, 
Jay Anderson\altaffilmark{4}, 
Harvey B. Richer\altaffilmark{5}, 
Ivan R. King\altaffilmark{6},
James P. Brewer\altaffilmark{5}, \\
Giovanni Carraro\altaffilmark{7},
Saul D. Davis\altaffilmark{5}, 
Gregory G. Fahlman\altaffilmark{8}, 
Brad M.~S. Hansen\altaffilmark{9}, \\
Jarrod R. Hurley\altaffilmark{10},
S\'ebastien L\'epine\altaffilmark{11},
David B. Reitzel\altaffilmark{9},
R. Michael Rich\altaffilmark{9}, \\
Michael M. Shara\altaffilmark{11},
Peter B. Stetson\altaffilmark{8}
}

\altaffiltext{1} {Based on observations with the NASA/ESA {\it Hubble Space Telescope}, obtained 
at the Space Telescope Science Institute, which is operated by the Association of Universities 
for Research in Astronomy, Inc., under NASA contract NAS5-26555.  These observations are 
associated with proposal GO-10424.}
\altaffiltext{2}{University of California Observatories/Lick Observatory, University of 
California at Santa Cruz, Santa Cruz, CA; jkalirai@ucolick.org}
\altaffiltext{3}{Hubble Fellow}
\altaffiltext{4}{Department of Physics and Astronomy, Rice University, Houston, TX}
\altaffiltext{5}{Department of Physics and Astronomy, University of British Columbia, 
Vancouver, BC, Canada}
\altaffiltext{6}{Department of Astronomy, University of Washington, Seattle, WA}
\altaffiltext{7}{Andes Fellow, Departamento de Astronom\'ia, Universidad de Chile, 
Santiago, Chile; and Department of Astronomy, Yale University, New Haven, CT}
\altaffiltext{8}{National Research Council of Canada, Herzberg Institute of Astrophysics, 
Victoria, BC, Canada}
\altaffiltext{9}{Department of Astronomy and Astrophysics, University of California at 
Los Angeles, Los Angeles, CA}
\altaffiltext{10}{Department of Mathematics and Statistics, Monash University, Clayton, Australia}
\altaffiltext{11}{Department of Astrophysics, American Museum of Natural History, New York, NY}

\begin{abstract}

As a by-product of high-precision, ultra-deep stellar photometry in the 
Galactic globular cluster NGC 6397 with the {\it Hubble Space Telescope}, 
we are able to measure a large population of background galaxies whose 
images are nearly point-like.  These provide an extragalactic reference 
frame of unprecedented accuracy, relative to which we measure the 
most accurate absolute proper motion ever determined for a globular cluster.  
We find $\mu_{\alpha}$\alphabcos = 3.56 $\pm$ 0.04~mas~yr$^{-1}$ and 
$\mu_{\delta}$ = $\rm-$17.34 $\pm$ 0.04~mas~yr$^{-1}$.  We note that the formal 
statistical errors quoted for the proper motion of NGC 6397 do not include 
possible unavoidable sources of systematic errors, such as cluster rotation.  
These are very unlikely to exceed a few percent.  We use this new proper motion 
to calculate NGC 6397's $UVW$ space velocity and its orbit around the 
Milky Way, and find that the cluster has made frequent passages through the 
Galactic disk.

\end{abstract}


\section{Introduction} \label{introduction}

Absolute proper motions of Galactic globular clusters are a necessity
for determining the orbits of these systems around the Milky Way.
These orbits are calculated by combining the proper motion of a cluster
with its radial velocity and distance from the Sun, and assuming a
Galactic potential.  The resulting space motion can constrain processes
of cluster origin and destruction as well as Galactic dynamics. 
By studying a large set of globulars with different demographics (e.g., 
metallicity), cluster subsystems can be identified and used to
understand formation scenarios of the various Milky Way components
(e.g., Dinescu, Girard, \& van Altena 1999; Dauphole et~al.\ 1996;
Searle \& Zinn 1978).

Historically, the transverse motions of Galactic globular clusters
have been determined by comparing positions of their stars on
photographic plates taken decades apart.  The resulting observed motions
are then converted to absolute proper motions by using a set of 
extragalactic objects to provide a zero-motion reference frame 
(e.g., Klemola, Jones, \& Hanson 1987), by using field stars whose 
absolute proper motions are known (e.g., Hanson et~al.\ 2004), or by 
using secular parallaxes of field stars (e.g., 
Cudworth 1979).  A summary of the kinematical studies of Galactic globular 
clusters using these different approaches is given in \cite{dinescu99}.  
For many of the clusters the proper motions have large errors, 
$\sim$1--2~mas~yr$^{-1}$, mainly because of the small number of 
reference objects and their low astrometric accuracy.  Magnitude/color 
dependent errors as well as aberrations near the edges of the photographic 
plates also contribute to the large errors in the absolute proper motions.

NGC 6397, one of the earliest-discovered globular star clusters 
(de Lacaille 1755), is at $\alpha_{\rm J2000}$ = 17:40:42.3, $\delta_{\rm J2000}$ =
$-$53:40:29.0 ($l$ = 338.2, $b$ = $-$11.96).  It is the second
nearest globular cluster to the Sun ($d \sim 2,600$~pc -- Gratton et~al.\ 2003), 
and one of the most metal-poor in the Milky Way, [Fe/H] = $-$2.0 \citep{gratton}.  
The previous wealth of data on the cluster has constrained most properties of 
NGC 6397, but the cluster's absolute proper motion has been measured only 
twice.  \cite{cudworth93} report the absolute proper motion of NGC 6397 to 
be $\mu_{\alpha}$\alphabcos = 3.30 $\pm$ 0.50 mas~yr$^{-1}$, 
$\mu_{\delta}$ = $\rm-$15.20 $\pm$ 0.60 mas~yr$^{-1}$.  The error bars in 
these measurements may have been underestimated; they relate to the 
comparison of the mean motion of field stars with the prediction from 
a model.  The second measurement of the proper motion of NGC~6397 
(Milone et~al.\ 2006) is based on a sample of $\sim$30 galaxies measured 
with the {\it Hubble Space Telescope} (\hst) WFPC2 and appeared during the 
writing of this Letter.  Their values are $\mu_{\alpha}$\alphabcos = 3.39 $\pm$ 
0.15 mas~yr$^{-1}$ and $\mu_{\delta}$ = $\rm-$17.55 $\pm$ 0.15 
mas~yr$^{-1}$, $\sim$15\% larger than the measurement reported by 
\cite{cudworth93} (see \S\,\ref{derivationmu}). 

In this Letter we present a direct measurement of the proper motion of 
NGC 6397 relative to a large sample of galaxies, using deep photometry, 
astrometry, and morphological distinction between stars and galaxies.

\section{The Data} \label{thedata}

\subsection{Image Analysis} \label{imageanalysis}

We imaged a single field of the globular star cluster NGC 6397 in March
and April of 2005 with the Advanced Camera for Surveys (ACS) on \hst\
(GO-10424).  The total allocated time was 126 orbits and the
observations were obtained in two broadband filters; 252 images in $F814W$, 
and 126 in $F606W$.  Several advances and refinements in 
data analysis were developed for the reduction of these data; these will be
fully presented in J.\ Anderson et~al.\ (2007, in preparation).  Some details
are also provided in Richer et~al.\ (2006).  Here we summarize the key
steps relevant to the present Letter.  We treated each local maximum in
any $F814W$ exposure as a potential detection, and collated all of the peaks.
We retained peaks that were found in nearly the same place in 90
images out of 252 (a 3$\sigma$ detection), and were the most significant 
ones within 7.5 pixels.  For {\it stellar} sources, we further required that 
the source meet two morphology criteria.  The first, CENRESID, is defined as the 
fractional flux remaining within the central 3x3 pixel region of a source 
after the best-fitting point-spread function (PSF) has been subtracted 
off.  The parameter is zero for stars that are well fit by the PSF.  The 
second criteria, ELONG, measures the elongation of each source and is 
zero for objects that have no excess flux over the PSF fit 
in any direction.  Finally, we eliminate false detections in the 
wings of bright stars and those caused by diffraction spikes.  Out of the 
$\sim$50,000 sources in our original catalog, 8,401 {\it stars} survive 
these cuts.  In addition, many extragalactic objects are found (see 
below).

\subsection{Measuring the Proper Motions} \label{measuringpropermotion}

Our {\it HST}/ACS pointing of NGC 6397 was specifically targeted to a field
$\sim5'$ southeast of the center of NGC 6397, which had been previously studied
with {\it HST}/WFPC2 in 1994 and 1997 \citep{king}.  The earlier observations 
cover about 60\% of our ACS field, so that we 
can measure the motions of a large fraction of our objects by comparing
their new positions with those at the previous epochs.  We first used
3,510 bright cluster stars to define a 6-parameter transformation from the 
frame of each WFPC2 archival image into our master frame.  Only cluster 
stars were used to define the transformations so there is no displacement 
between the archival position and the 2006 position (i.e., our zero-point 
of motion corresponds to the systemic motion of the cluster).  Since a 
field star may happen to lie along the cluster main-sequence, we 
iteratively rejected any star with a transformation residual $>$0.25 
pixels in the master frame.

For each of the 50,000 detections we used the transformations to determine 
where the star would be in each of the archival exposures.  Since we do not 
{\it a priori} know the proper motion of any object that we measured on 
the ACS images, we examined all of the WFPC2 detections in the vicinity of 
the object to see which of them might correspond to the ACS detection.  To 
compare the 1994 and 1997 archival data on the same footing, we converted 
each possible WFPC2 detection into an estimate for the 10-year displacement.  
We then examined all of these and chose those with the same implied proper 
motion as successful identifications.  The errors in the proper motions are 
calculated as the dispersion in the proper motion measurements from the multiple 
first-epoch images, divided by the square root of the number of first epoch 
images used.

\subsection{An Extragalactic Zero Motion Frame of Reference}
\label{extragalacticframe}


\begin{figure}
\epsscale{1.15} 
\plotone{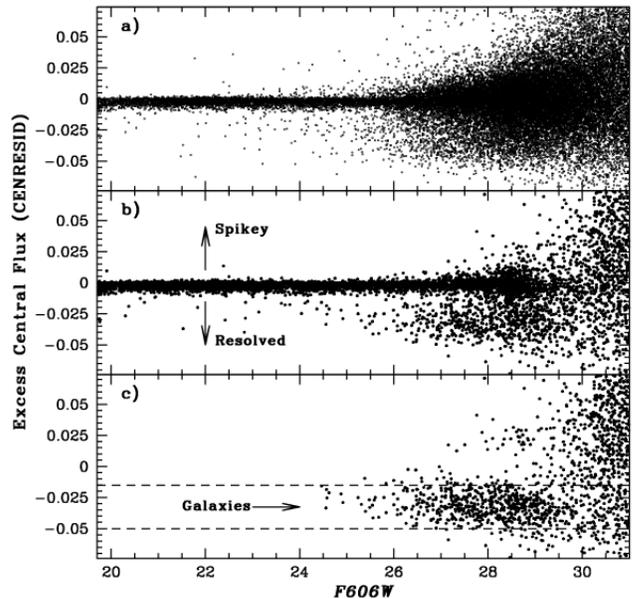} 
\caption{The fractional 
excess central flux defined by CENRESID (see \S\,\ref{imageanalysis}) 
is used to separate stars from galaxies as discussed in 
\S\,\ref{extragalacticframe}.
\label{fig:sharp}}
\end{figure}


The most important step in directly calculating an accurate proper motion for 
NGC 6397 is the choice of reference objects, which need
to be extragalactic but as point-like as possible.  For this we use the
morphological parameters defined earlier.  The top part of Figure~1a shows 
the CENRESID parameter for all sources in our data, as a function of $F606W$ 
magnitude.  Most objects (the stars) form a tight band with CENRESID 
$\sim$ 0.  In the middle panel 
(Figure~1b), we isolate those objects that pass a cut to eliminate false 
detections around bright stars and intersections of diffraction spikes, and for 
which we have measured a proper motion (see section \ref{measuringpropermotion}).  
A clean sequence of stars is seen extending nearly to $F606W$ = 30.  A second clump 
of sources with negative CENRESID values is also clearly seen.  Negative values 
imply that these objects have less flux at their centers than the PSF would 
predict, making them strong candidate galaxies.  In Figure~1c we isolate these 
galaxies by eliminating stellar objects.  The dashed lines show the boundaries 
of our galaxy sample, $-$0.05 $<$ CENRESID $<$ $-$0.015.

This sample, selected entirely on the basis of morphology, contains 398 galaxies (one 
object is a clear outlier at the bright end and was therefore removed as it likely 
represents a field star interloper).  For the best of these galaxies (bright, 
compact, and isolated), we can easily measure the proper motion to within 
0.25~mas~yr$^{-1}$.  This is the first time that such a large 
sample of almost point-like galaxies has been used to measure proper motions.  

\section{Derivation of the Proper Motion and the Tangential Velocity of 
NGC 6397} \label{derivationmu}

To determine the proper motion of NGC 6397 we need to measure the difference in 
proper motion of the galaxy distribution determined above and the cluster distribution.  
This involves two steps.  First, we convert our observed $X$ and $Y$ ACS pixel 
motions (measured relative to NGC 6397 stars) into equatorial coordinates 
($\alpha$, $\delta$) by rotating the field into the correct orientation.  We also 
multiply the motions by the pixel scale (0\farcs05) and normalize to units of 
milli arcseconds per year.  Second, we center the mean of the 398 galaxies in 
the galaxy distribution to (0, 0) on this new plane with a simple 
translation.


\begin{figure}
\epsscale{1.15} 
\plotone{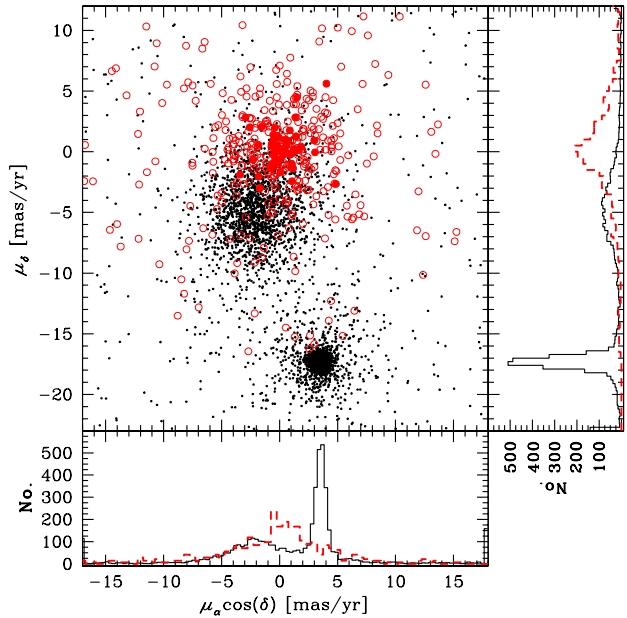} 
\caption{Proper-motion
diagram for objects along the NGC 6397 line of sight.  The tight clump
near the bottom of the diagram represents cluster members, the more
diffuse clump near the center of the diagram consists of field stars
along the line of sight, and the larger open circles represent
galaxies (see \S\,\ref{extragalacticframe}).  The 50 galaxies that provide 
the most weight to our measurement are shown as large filled circles.  The 
histograms on the bottom and right better illustrate the three different 
populations in the proper-motion diagram.  For clarity, the galaxy numbers 
in these two panels ({\it dashed curve}) have been scaled up by a factor of 7. 
\label{fig:pmd}}
\end{figure}


Figure~2 presents our final proper-motion diagram for this field.  The
components are expressed in right ascension and declination, such that the
total proper motion $\mu$ = $\sqrt{(\mu_{\alpha}\testalphabcos)^{2} + (\mu_{\delta})^{2}}$.  
The small dots indicate all objects that pass the criteria in \S\,\ref{imageanalysis}.  
These clearly form two clumps in the proper motion diagram, the cluster NGC
6397 being represented with the tighter clump near the bottom of the
diagram and the bulk of the field stars being concentrated in the more
diffuse clump near the center of the diagram.  The morphologically selected 
galaxies, shown as larger circles, also clearly clump together in this plane and 
have a different centroid of motion from that of the field stars.  We find 
the centroid of the proper motions of these galaxies by taking a
weighted mean, with weights $w_i$ = 1/$(\sigma_{i})^2$ where the
$\sigma_i$ are the estimated errors of the proper motions of the
individual objects.  Although the proper motions of many galaxies go
into this mean, it is in fact dominated by those with the highest 
weight.  The 50 galaxies that contribute the most to our extragalactic 
reference frame are distinguished as filled circles in Figure~2.

Relative to this standard of zero motion, the absolute proper motion of
NGC 6397 is $\mu_{\alpha}$\alphabcos = 3.56 $\pm$ 0.04 mas~yr$^{-1}$ and 
$\mu_{\delta}$ = $\rm-$17.34 $\pm$ 0.04 mas~yr$^{-1}$.  This is the most 
accurate measurement of the absolute proper motion of a
globular cluster, to date.  We note that our field is far removed from 
the cluster center and possible cluster rotation therefore injects some 
additional systematic uncertainty.  Given the 3.5~km~s$^{-1}$ dispersion 
of radial velocities \citep{pryor}, the {\it maximum} observed rotational 
velocity is likely to be less than 2~km~s$^{-1}$ for an axial ratio of 0.93 for 
the cluster (White \& Shawl 1987).  However, we do not know what 
the inclination is.  If we are observing the cluster pole-on, then at a point 
removed from the cluster center, rotation would be across the line of sight.  
Considering the ellipticity, it is unlikely that this is the case.  If our 
line of sight were perpendicular to the axis of rotation, every line of sight 
would have zero mean transverse velocity, because symmetry leads to a 
cancellation between motions in front of and behind the midpoint.  Considering 
these effects, which are difficult to estimate, rotation might introduce an 
uncertainty of some modest fraction of a kilometer per second, or an amount 
comparable to our measuring error.  Future radial velocity and proper motion 
measurements over a large radial distance in the cluster may permit a derivation 
of this cluster rotation (see e.g., van~de~Ven et~al.\ 2006).

Our measurement of the absolute proper motion of NGC 6397 is $\sim$14\% larger 
than the \cite{cudworth93} value, but is in good agreement with the
recent measurement by \cite{milone}, within their fourfold larger error 
bounds.  For completeness, the proper motion of NGC 6397 in Galactic 
coordinates is $\mu_{l}$cos($b$) = $\rm-$13.27 $\pm$ 0.04~mas~yr$^{-1}$ 
and $\mu_{b}$ = $\rm-$11.71 $\pm$ 0.04~mas~yr$^{-1}$.

It is important to understand why our error is only a quarter as large as
that of Milone et~al., and also significantly better than other similar 
studies (e.g., the Kalirai et al.\ 2004 study of the globular cluster M4).  The 
comparison does not reflect in any way on the quality of the 
measurements in these other studies, which represent the state of the art 
in astrometry of images of objects that look like galaxies.  The
difference is that we have found a method for morphological selection of
a large sample of point-like galaxies that mere eye examination cannot
distinguish from stars.  The best of these objects are incomparably
better for astrometry than even the sharpest-centered galaxies that can
be recognized by eye.  The measurement by Milone et~al. is limited to 
only 33 extragalactic sources, many of which have a large dispersion 
and are therefore down-weighted.  Their final result is derived almost 
entirely from a few of the brightest sources with the sharpest nuclei.

With our estimated distance of 2,600 $\pm$ 130~pc to NGC 6397, we find its 
tangential velocity to be $v_{\alpha}$ = 43.3 $\pm$ 2.2~km~s$^{-1}$ and 
$v_{\delta}$ = $-$211.3 $\pm$ 10.6 km~s$^{-1}$.  The error budget in the 
tangential velocity of NGC 6397 is dominated entirely by the error in the 
distance, which we estimate to be $\sim$5\%.  (The proper motions alone would 
contribute an error of $<$0.5 km~s$^{-1}$.)  This distance and corresponding 
uncertainty comes from a fit of the subdwarfs in \cite{reid98} to the new, 
very tightly constrained main sequence of NGC 6397 in Richer et~al.\ (2006).  
The subdwarf sample has accurate parallax measurements (to within 10\%), 
and shifting the mean distance by any more than 5\% would cause a large 
discrepancy between these two loci that cannot be recovered by mild 
changes in, for example, metallicity.  The third component of the cluster's 
motion, its radial velocity, was determined by \cite{milone} to be 18.36 $\pm$ 
0.13~km~s$^{-1}$.  

To help interpret the velocity of NGC 6397, we now calculate the $UVW$ 
motion of the cluster, where the $U$ component is positive toward the Galactic 
center, $V$ is positive in the direction of Galactic rotation, and $W$ is 
positive toward the north Galactic pole (see e.g., Johnson \& Soderblom 1987).  
After correcting for a Solar motion of ($U_\odot$, $V_\odot$, $W_\odot$)
= ($+$10.00 $\pm$ 0.36, $+$5.25 $\pm$ 0.62, $+$7.17 $\pm$ 0.38) km~s$^{-1}$ 
\citep{dehnen}, we find ($U$, $V$, $W$) = ($\rm-$61.2 $\pm$ 4.4, $\rm-$140.5 
$\pm$ 7.0, $\rm-$136.3 $\pm$ 7.0) km~s$^{-1}$.  The same 5\% distance uncertainty 
has been included in the error budget on the space motion, and again, entirely 
dominates the error.  Relative to the Galactic center, the ($\Pi$, $\Theta$) 
components of NGC 6397 are ($\rm-$46.6 $\pm$ 4.4, 93.3 $\pm$ 7.0) km~s$^{-1}$, 
assuming $R_0$ = 8~kpc and $\Theta_0$ = 225~km~s$^{-1}$.


\begin{figure}
\epsscale{1.18}
\plotone{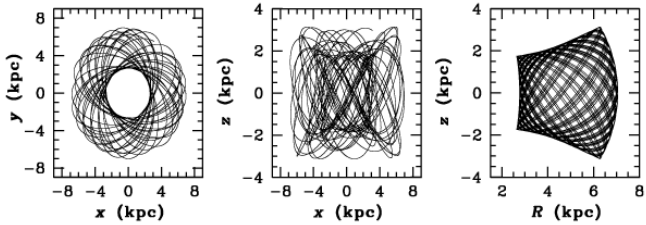}
\caption{Orbit of NGC 6397 shown in the disk plane ({\it left}) and
perpendicular to the disk ({\it middle and right}).  The orbit shows that NGC
6397 has crossed the Galactic disk many times.  The pericentric radius
of the cluster orbit is found to be $R_p \sim$ 2.6~kpc and the apocentric
radius is $R_a \sim$ 7.0~kpc.  Further details on the cluster orbit are
provided in \S\,\ref{orbitanalysis}.
\label{fig:orbit}}
\end{figure}


\section{The Orbit of NGC 6397} \label{orbitanalysis}

To integrate the Galactic orbital motion of NGC 6397, we use the initial conditions
derived above and a Galactic gravitational potential with the bulge and halo 
represented as two spherical components and the disk represented as an axisymmetric 
component (see Dauphole \& Colin 1995 for details).  Using a Runge-Kutta 
fourth-order integrator with time steps of 0.1 Myr, we integrate the orbit for 5~Gyr.  
In Figure~3 we show NGC 6397's orbit in the disk plane (left) and perpendicular to the disk
(middle and right).  Clearly, the cluster's orbit is well confined within a 
relatively small galactocentric radius (as noted by Dauphole et~al.\ 1996 and 
Milone et~al.\ 2006).  NGC 6397 orbits our Galaxy in the direction of Galactic 
rotation.  The cluster has spent most of its time close to the Galactic 
disk, never reaching a height of more than $\sim$3.1~kpc above the plane.  
The orbit crosses the plane many times and therefore the cluster has likely 
experienced shocks from the Galactic disk.  The timescale for this is less 
than 100~Myr, shorter than the half-mass relaxation time (300 Myr - Harris 1996). 
The apocentric radius, $R_a$, is found to be $\sim$7.0~kpc, and 
therefore the cluster has not made any large excursions into the Galactic halo
(eccentricity, ($R_a - R_p$)/($R_a + R_p$) = 0.46 $\pm$ 0.02).  The cluster may 
have interacted with the Galactic bulge and suffered shocks from pericentric 
passages ($R_p$ $\sim$2.6~kpc).  Given the cluster's present location 
($R_{GC}$ = 5.7~kpc) and velocity, we can conclude that it is approaching 
its apocenter.  

We also tested the orbital motion assuming different starting conditions, 
$R_0$ = 7.5--8.5~kpc, and find that the cluster's orbit is similar to 
that computed above (the pericenter radii shifts by $\sim$5\%).  Although 
the orbit is more sensitive to the adopted distance to NGC 6397 (e.g., 
see Milone et~al.\ 2006), a 5\% distance uncertainty does not have a major 
affect on the apocenter and pericenter radii.  Finally, we note that both the 
orbital acceleration and the perspective acceleration (caused by our line 
of sight intersecting a point slightly removed from the center of 
NGC 6397 - Schlesinger 1917) of the cluster are negligible 
(several hundred times smaller than our error bars).

\section{Conclusions}

The results of this Letter are a consequence of our having high-precision
photometry, including morphological indices that allow the selection of
star-like objects that are actually galaxies.  The use of these as a
reference standard has allowed us to measure the absolute proper motion
of NGC 6397 with unprecedented accuracy and to compute a more reliable 
orbit for it.  We find that the cluster has interacted with
the Galactic disk and bulge many times over the past 12~Gyr. Such tidal
interactions may modify the cluster mass function and careful determination 
of this function coupled with dynamical models could yield evidence of this 
interaction (Dinescu, Girard, \& van Altena 1999). 

\acknowledgements
We would like to thank A.~R. Klemola, R.~B. Hanson, and L.~R. Bedin for 
discussions regarding the conversion of proper motions into space velocities 
and corresponding error analysis. JSK is supported by NASA through Hubble 
Fellowship grant HF-01185.01-A, awarded by the Space Telescope Science Institute 
(STScI), which is operated by the Association of Universities for Research in
Astronomy, Inc., under NASA contract NAS5-26555. Support for
this work was also provided by grant HST-GO-10424 from NASA/STScI, the Natural 
Sciences and Engineering Research Council of Canada, and the Canada-US Fulbright 
Program.

\end{document}